\documentclass[]{sn-jnl}
\usepackage{graphicx}%
\usepackage{multirow}%
\usepackage{amsmath,amssymb,amsfonts}%
\usepackage{amsthm}%
\usepackage{mathrsfs}%
\usepackage[title]{appendix}%
\usepackage{xcolor}%
\usepackage{textcomp}%
\usepackage{manyfoot}%
\usepackage{booktabs}%
\usepackage{algorithm}%
\usepackage{algorithmicx}%
\usepackage{algpseudocode}%
\usepackage{listings}%
\usepackage{ulem}
\usepackage{xspace}
\usepackage{comment}
\usepackage{natbib}

\theoremstyle{thmstyleone}%
\theoremstyle{thmstyletwo}%

\theoremstyle{thmstylethree}%

\newcommand{\ben}{\begin{enumerate}}
\newcommand{\een}{\end{enumerate}}
\newcommand{\Fermi}{\textit{Fermi}\xspace}
\newcommand{\persec}{$\,{\rm s}^{-1}$\xspace}

\usepackage[utf8]{inputenc}

\begin{document}

\title[SMBH and MeV emission]{Supermassive black holes and their surroundings: MeV signatures}


\author*[1]{\fnm{Tullia} \sur{Sbarrato}}\email{tullia.sbarrato@inaf.it}
\author[2]{\fnm{Marco} \sur{Ajello}}
\author[3,4]{\fnm{Sara} \sur{Buson}}
\author[5]{\fnm{Denys} \sur{Malyshev}}
\author*[6]{\fnm{Dmitry V.} \sur{Malyshev}}\email{dmitry.malyshev@fau.de}
\author[7]{\fnm{Reshmi} \sur{Mukherjee}}
\author[1]{\fnm{Gianpiero} \sur{Tagliaferri}}
\author[1]{\fnm{Fabrizio} \sur{Tavecchio}}


\affil[1]{\small INAF - Osservatorio Astronomico di Brera, via E.\ Bianchi 46, 23807 Merate (LC), Italy}
\affil[2]{\small Department of Physics and Astronomy, Clemson University, 29631 Clemson, SC, USA}
\affil[3]{\small Fakultät für Physik und Astronomie, Julius-Maximilians-Universität Würzburg, Emil-Fischer-St. 31, D-97074, Würzburg, Germany}
\affil[4]{\small Deutsches Elektronen-Synchrotron DESY, Platanenallee 6, 15738 Zeuthen, Germany}
\affil[5]{\small Institut für Astronomie und Astrophysik Tübingen, Universität Tübingen, Sand 1, 72076 Tübingen, Germany}
\affil[6]{\small Erlangen Centre for Astroparticle Physics, Nikolaus-Fiebiger-Str.\ 2, 91058 Erlangen, Germany}
\affil[7]{\small Department of Physics and Astronomy, Barnard College, 3009 Broadway, 10027 New York, NY, USA}


\abstract{
The gravitational potential of supermassive black holes is so powerful that it triggers some of the most intense phenomena in the Universe.
Accretion onto these objects and relativistic jet emission from their vicinity are observable across a wide range of frequencies and throughout cosmic history. However, despite this wealth of data, many aspects of their underlying mechanisms remain elusive.
Investigating this phenomena across all frequencies is crucial, yet some energy windows are still poorly explored. 
One such window is the MeV energy range: many key signatures related to the emission from the SMBH environment -- both in quiescent and active phases -- are expected to lie between one and several hundreds MeV.
In this work, we explore some of the open questions regarding the behavior and emission processes in the surroundings of SMBHs, and how these questions might be approached. From the elusive nature of {\it Fermi} bubbles around our Galactic Centre, to the origin of high-energy neutrinos in the nuclei and jets of Active Galactic Nuclei, to the nature and emission mechanisms of the most powerful blazars, the MeV window stands out as a crucial key to  understanding SMBH physics.
}


\maketitle

\section{Introduction}

The existence of supermassive black holes (SMBHs) at the centre of massive galaxies have been first suggested by \cite{salpeter64} and \cite{lynden-bell69}.
The main argument in support of their presence is the extreme brightness of some galactic nuclei, that can exceed the total host galaxy emission. 
Such intense brightness could not be justified just by nuclear processes, involved in stellar evolution, and thus accretion processes on very massive compact objects were introduced. 
Depending on the assumed physical conditions, in fact, accretion can release between 5\% and $\sim30$\% of the rest mass-energy of the material infalling toward the compact object \citep{thorne74}, allowing for large amount of radiation, comparable with that observed in bright nuclei. 

In the following decades, other convincing evidences of the existence of SMBHs in galactic centres were proposed, even by focusing on our own Galactic centre \citep{kormendy95}. 
Radio observations of SgrA$^*$ \citep{balick74}, in fact, were quickly associated with a SMBH \citep{falcke93}. 
Finally, continuous monitoring of the Galactic centre with very high resolution NIR imaging revealed a population of high proper motion stars moving in an extremely massive gravitational potential \citep{ghez98}. This ultimately confirmed that a $\sim4$  million solar masses SMBH lived at the centre of our Galaxy. 
The advent of high-energy telescopes, and {\it Fermi} in particular, allowed to open a new window on SgrA$^*$ current and past activity. 
Extended, prominent $\gamma$-ray emitting structures dubbed as {\it Fermi} bubbles were discovered emerging from the Galactic centre \citep{2010ApJ...724.1044S_Su_bubbles}, 
followed by the detection of more extended soft-X-ray bubbles by eROSITA \citep{predehl20},
suggesting past outflows (or maybe supernovae activity, see Section \ref{sec:fermi-bubbles}). 

When accretion of matter on the SMBH is significant, the galaxy hosts an Active Galactic Nucleus \citep[AGN,][]{urry95}. 
This is thought to happen in just $1-10\%$ galaxies, that is curious since excess nuclear brightness was the very first hint toward the presence of SMBHs in the centre of galaxies. AGN emission can cover the whole electromagnetic spectrum, spanning few orders of magnitudes up to $L\sim10^{48-49}$erg/s in the MeV--$\gamma$-ray bands, depending on the accretion powers involved and the specific AGN features.
In about $10\%$ AGN, bipolar relativistic jets composed of ionised plasma are emitted in the vicinity of the SMBH, extending up to thousands of parsecs, often on scales larger than their host galaxies. Strong radio emission typically characterizes these structures, both in their collimated, acccelerated sections, and in the most extended parts. 
When impacting with the intergalactic medium, indeed, the magnetic field permeating the environment and the particle population injected from the jet direction are responsible for strong, isotropic synchrotron emission.
When relativistic jets are aligned close to our line-of-sight, AGN emit up to very high-energies thanks to their strong relativistic beaming These sources are classified as blazars. 
In the X-ray and $\gamma$-ray bands, these sources represent a very significant population. 
This has immediately been clear when the first high-energy satellite started to take data.
The extragalactic $\gamma$-ray sky was and is dominated by blazars, starting from the first Energetic Gamma-Ray Experiment Telescope (EGRET) Source and blazar Catalogs  \citep[][]{Mukherjee97,fichtel94}, the First Catalog of Active Galactic Nuclei Detected by the {\it Fermi} Large Area Telescope  \citep[LAT, 1LAC,][]{abdo10}, and the first {\it Swift} Burst Alert Telescope (BAT) survey of AGN \citep{tueller08}.
Particularly the results obtained by {\it Swift}/BAT made clear that the extragalactic soft $\gamma$-ray background (i.e.\ $>500$ MeV) is likely dominated by the integrated emission from all the existing jetted AGN \citep{ajello09}.
This strikes the importance of understanding the outflow and relativistic jet emission from the vicinity of active SMBH.

While the general picture of SMBHs in galactic nuclei is pretty clear, with the possibility of different matter accretion rates, and a wide range of possible outflows, many details are still very obscure. 
From the lowest accretion regimes, maybe linked to extended relic structures such as the {\it Fermi} bubbles observed around SgrA$^*$, up to the fastest and most efficient ones, linked to very powerful relativistic jets visible up to high-redshift, the particle composition, cosmological evolution, launching and acceleration mechanisms are still some of the most interesting open questions of AGN physics.

All these processes are being monitored across the whole electromagnetic spectrum, but the energy coverage is not uniform. 
A significant lack in our observing capabilities and facilities lies between the hardest X-ray frequencies and the $\gamma$-ray regime: the MeV energy range has been historically less sampled, with few instruments covering it only partially, or with limited sensitivity. 
In this chapter, we are exploring how MeV observations could be key for advancing our knowledge of the SMBH nearest surroundings, both in terms of their emission and the physics connected.

\section{Accretion and outflows: the extremes}

\subsection{SgrA* and {\Fermi} Bubbles}
\label{sec:fermi-bubbles}

One of the indirect evidences of the past activity at or near the SMBH Sgr A* is the existence of large gamma-ray emitting lobes above and below the Galactic Centre (GC), the \Fermi bubbles (FBs)~\citep{2010ApJ...724.1044S_Su_bubbles, 2014ApJ...793...64A_FB_Fermi}. 
FBs were discovered in the \Fermi-LAT gamma-ray data in 2010~\citep{2010ApJ...724.1044S_Su_bubbles}.
They have a size of about 10 kpc above and below the GC, which is similar to kpc scale radio lobes observed in Seyfert 
galaxies~\citep{1993ApJ...419..553B_Seyferts_winds, 2006AJ....132..546G_Seyferts_outflow_review}.
The origin of the FBs, either due to an outflow from Sgr A*~\citep{2011MNRAS.415L..21Z_Zobovas_burst_model, 2012ApJ...756..181G_Guo_AGN,
2013MNRAS.436.2734Y_Karen_Yang}, a starburst activity near the GC~\citep{2014MNRAS.444L..39L}, or a quasi-stationary injection of power from mini-bursts of Sgr A*~\citep{2011ApJ731L17C_FB_bursts, 2017MNRAS.467.3544S_Sarkar_OVIII-OVII, 2023ApJ...951...36S_Sarkar_minijets} or supernovae near the GC~\citep{2011PhRvL.106j1102C_Crocker-Aharonian_FB_Model, 2015ApJ...808..107C}, is still a highly debated topic.
In the AGN / starburst hypothesis the FBs are inflated on the timescale of millions to tens of millions of years. In this case the leptonic origin of the gamma-ray emission is preferred, provided that the density of gas at high latitudes is not sufficient for the hadronic production of gamma rays.
In the quasi-stationary hypothesis, 
a sufficient accumulation of hadronic cosmic rays (CRs) for the gamma-ray production at high latitudes can be achieved over the timescales of hundreds of millions to billions of years~\citep{2011PhRvL.106j1102C_Crocker-Aharonian_FB_Model, 2015ApJ...808..107C}.
Understanding the nature of the gamma-ray emission from the FBs (leptonic vs hadronic) can help to separate the different models of the FBs, but the nature of the gamma-ray emission is also an open question.
For a recent review of FBs see, e.g., \cite{2024A&ARv..32....1S_Sarkar_FB_review}.

A possible avenue for understanding the origin of FBs is the gamma-ray observations near the base of FBs~\citep{2017ApJ...840...43A_LAT_GCE_FB_base, 2019A&A...625A.110H_LLB_Herold_Malyshev}.
These observations suggest that emission near the base of the FBs has a brighter intensity and extends to higher energies compared to gamma-ray emission from FBs at latitudes $|b| > 10^\circ$.
Also at the base of FBs, both leptonic and hadronic models can explain the gamma-ray emission \citep{2019A&A...625A.110H_LLB_Herold_Malyshev}. Although there is a sign of asymmetry of the gamma-ray emission at the FBs base, which is more natural for a starburst explanation of the FBs origin, the asymmetry can also be explained in the AGN scenario by the orientation of the jet from Sgr A* or by interaction of the spherical outflow from Sgr A* with the dense molecular clouds.

\begin{figure}[h]%
\centering
\includegraphics[width=0.8\textwidth]{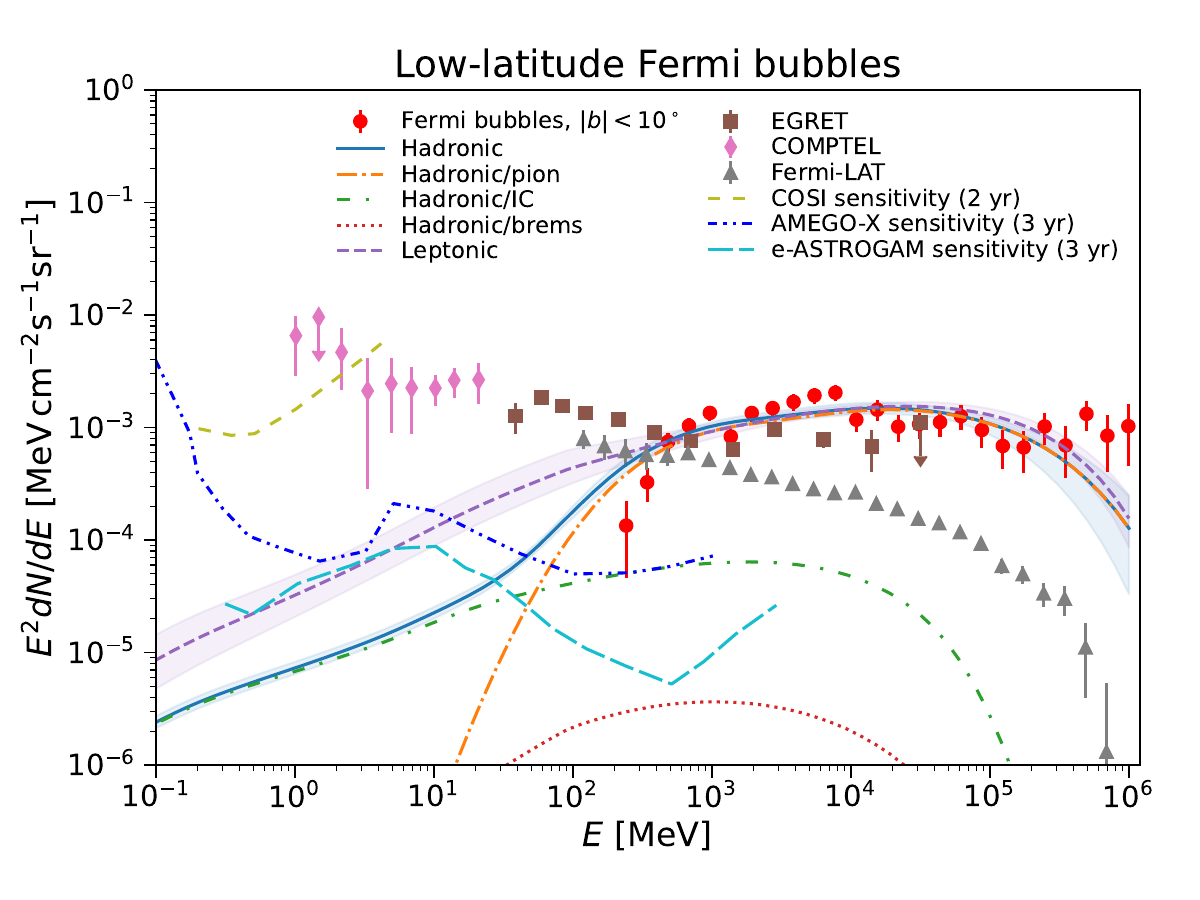}
\caption{Intensity of emission at the base of FBs within $|b| < 10^\circ$ 
\citep[red circles, ][]{2017ApJ...840...43A_LAT_GCE_FB_base}. 
Blue solid line shows the hadronic model of the gamma-ray emission. Dashed orange line, green sparse dash-dotted line, and dotted red lines show the primary $\pi^0$, secondary IC, and secondary bremsstrahlung components in the hadronic scenario respectively. Dashed purple line - leptonic scenario of gamma-ray emission (dominated by IC emission). Bands show the 1 sigma model uncertainty ranges for statistical plus 10\% systematic uncertainties in the data. Pink diamonds, brown squares, and grey upward triangles show the extragalactic diffuse gamma-ray background measured by 
COMPTEL~\citep{2000AIPC..510..467W},
EGRET~\citep{2004ApJ...613..956S_EGRET_IGRB}, and 
\Fermi-LAT~\citep{2015ApJ...799...86A_IGRB_Fermi} respectively.
Yellow sparse dashed line shows expected COSI sensitivity after 2 years of observations~\citep{2023arXiv230812362T} for an extended source with the area of the low-latitude FBs $\Omega \approx 0.14$~sr.
Expected AMEGO-X~\citep{2022JATIS...8d4003C} 
and e-ASTROGAM~\citep{2018JHEAp..19....1D}
sensitivities for low-latitude FBs after 3 years of observations are shown by blue dash-dot-dotted and cyan long-dashed lines respectively.
}
\label{fig:FB_spectrum}
\end{figure}

\begin{figure}[h]%
\centering
\includegraphics[width=0.8\textwidth]{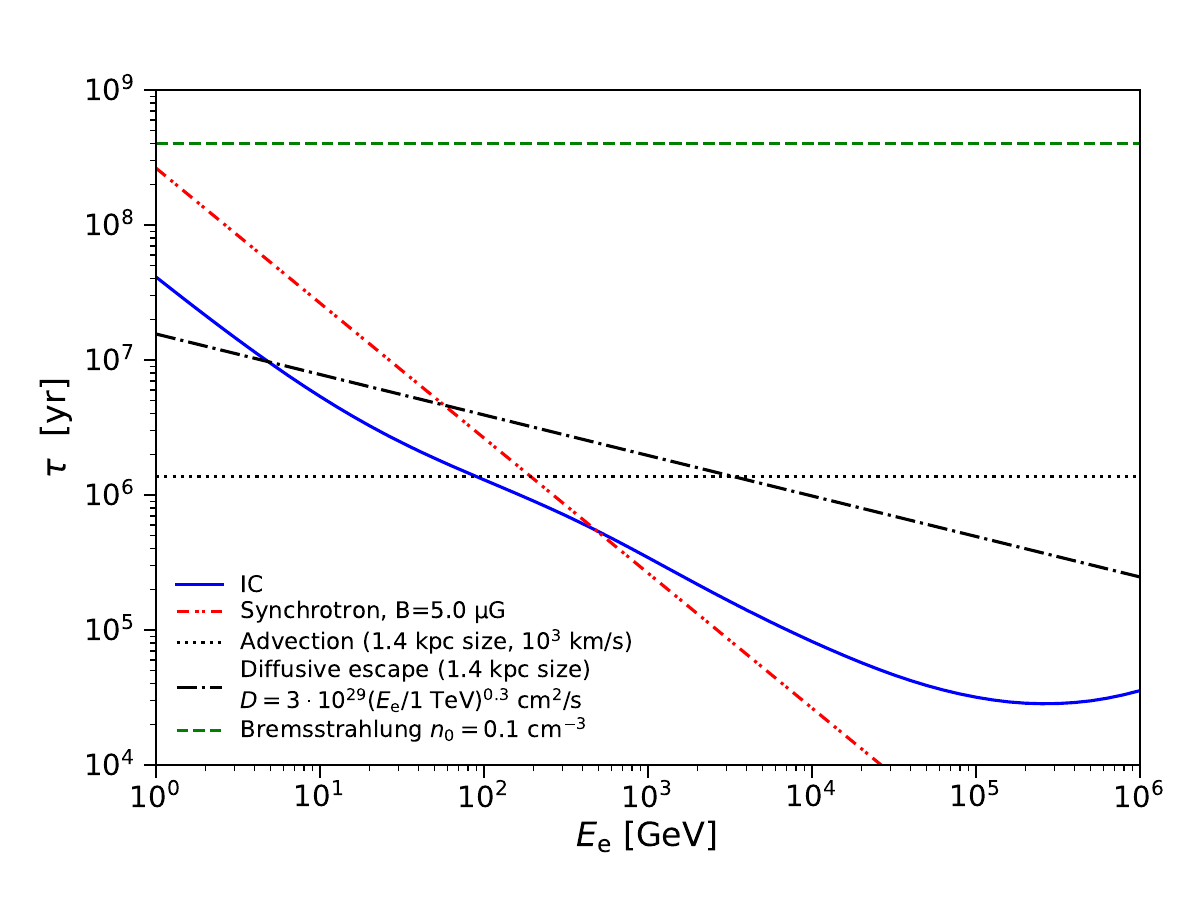}
\caption{Escape and cooling times for the CR electrons at the base of the FBs ($|b| < 10^\circ$). The advective and diffusive escape times are also relevant for the CR protons. For the calculation of IC losses we take volume averaged interstellar radiation fields of \cite{2017ApJ...846...67P} based on \cite{1998ApJ...492..495F}.
For the diffusion, we assume a similar diffusion coefficient as the local one~\citep{2012ApJ...752...68V_diffusion}.}
\label{fig:tcool}
\end{figure}

Observations of MeV $\gamma$-rays near the GC can help to constrain the nature of the gamma-ray emission and the models of the FBs.
Figure~\ref{fig:FB_spectrum} shows
the intensity of emission at the FBs base (i.e., for $|b| < 10^\circ$) as red circles~\citep{2017ApJ...840...43A_LAT_GCE_FB_base}.
The hadronic model of emission at the base of FBs is shown by the blue solid line. 
The spectrum of CRp is modeled by a power-law with an exponential cutoff function, which we fit to the spectral points of gamma-ray emission at the FB base. 
The band around the model shows the 1 sigma deviations of the model due to statistical uncertainty in the data plus 10\% systematic uncertainty added in quadrature to statistical uncertainty to allow for systematic uncertainties in the modeling and instrument response functions.
The spectrum of gamma rays created in the interactions of CRp with gas is shown by the orange dash-dotted line \citep[we assume the gas density of 0.1~cm$^{-3}$,][]{2007A&A...467..611F_Ferriere}.
We take into account the IC emission from the secondary electrons and positrons created in the hadronic interactions (shown by the green sparse dash-dotted line).
The corresponding bremsstrahlung emission due interactions of secondary $e^\pm$ with gas is shown by the red dash-dotted line.
In the leptonic model, we assume a power-law with exponential cutoff spectrum of CR electrons.
The inverse Compton (IC) emission in the leptonic model is shown by the purple dashed line, the bremsstrahlung emission is subdominant to the IC component (similarly to the IC and bremsstrahlung components in the hadronic scenario) and is not shown on the plot.
In the calculation of the flux from the secondary leptons in the hadronic scenario we take into account energy losses due to IC, bremsstrahlung, and synchrotron emission as well as escape due to diffusion \citep[assuming a diffusion coefficient similar to the local one;][]{2012ApJ...752...68V_diffusion} and outflow with a velocity of 1000 km\persec~\citep{2017ApJ...834..191B_outflow_velocity}.
The corresponding timescales are shown in Figure~\ref{fig:tcool}.
The most important energy loss mechanism below about 100 GeV is the advective escape. 
At a few hundred GeV the timescales for the advective escape, IC, and synchrotron losses are similar to each other. 
At higher energies the losses are dominated by the synchrotron emission.
We find that the gamma-ray flux in leptonic scenario is predicted to be a factor of 5 larger than the gamma-ray flux in hadronic scenario for $E \lesssim 50$~MeV.
The reason for this drop is the cutoff at energies below 100 MeV of the primary gamma-ray production and advective escape of secondary electrons. 
For much lower escape velocities than 1000 km\persec, e.g., for velocities comparable to the speed of sound in the ionized hydrogen plasma above the GC, the IC emission is a dominant energy loss mechanism between a few GeV and a few hundred GeV. 
In this case, the secondary IC production at tens of MeV would be at a comparable level to the IC emission in the leptonic scenario.
The sensitivities for 
COSI~\citep{2023arXiv230812362T}, 
AMEGO-X~\citep{2020SPIE11444E..31K}, and
e-ASTROGAM~\citep{2018JHEAp..19....1D}
to detect an extended source at 3 sigma level with the solid angle of the FBs at low latitudes are shown as yellow sparse dashed line, blue dash-dot-dotted line, and cyan long-dashed line, respectively.
We assume 2 years of observations for COSI and 3 years of observations for e-ASTROGAM and AMEGO-X.
The sensitivity in intensity units is computed by dividing the point-source flux sensitivity with $\sqrt{\Omega_{PS} \Omega_{\rm ROI}}$, where $\Omega_{\rm ROI} = 0.14$~sr is the solid angle of the low-latitude FBs and $\Omega_{PS}$ is an effective area of emission from a point-like source (PS). For e-ASTROGAM and AMEGO-X we take $\Omega_{PS} = \pi R_{68\%}^2$ with the 68\% containment radii at 50 MeV of $2.7^\circ$ and $5^\circ$, respectively. For COSI we use half-width half-maximum radius of $2^\circ$.
We find that the difference in the gamma-ray emission at the base of the FBs should be possible to detect with the next generation MeV instruments such as e-ASTROGAM and AMEGO-X.

\subsection{X-ray Corona and outflow}
\label{subsect:seyferts}

\begin{figure*}[!ht]
    \centering
    \includegraphics[scale=0.4]{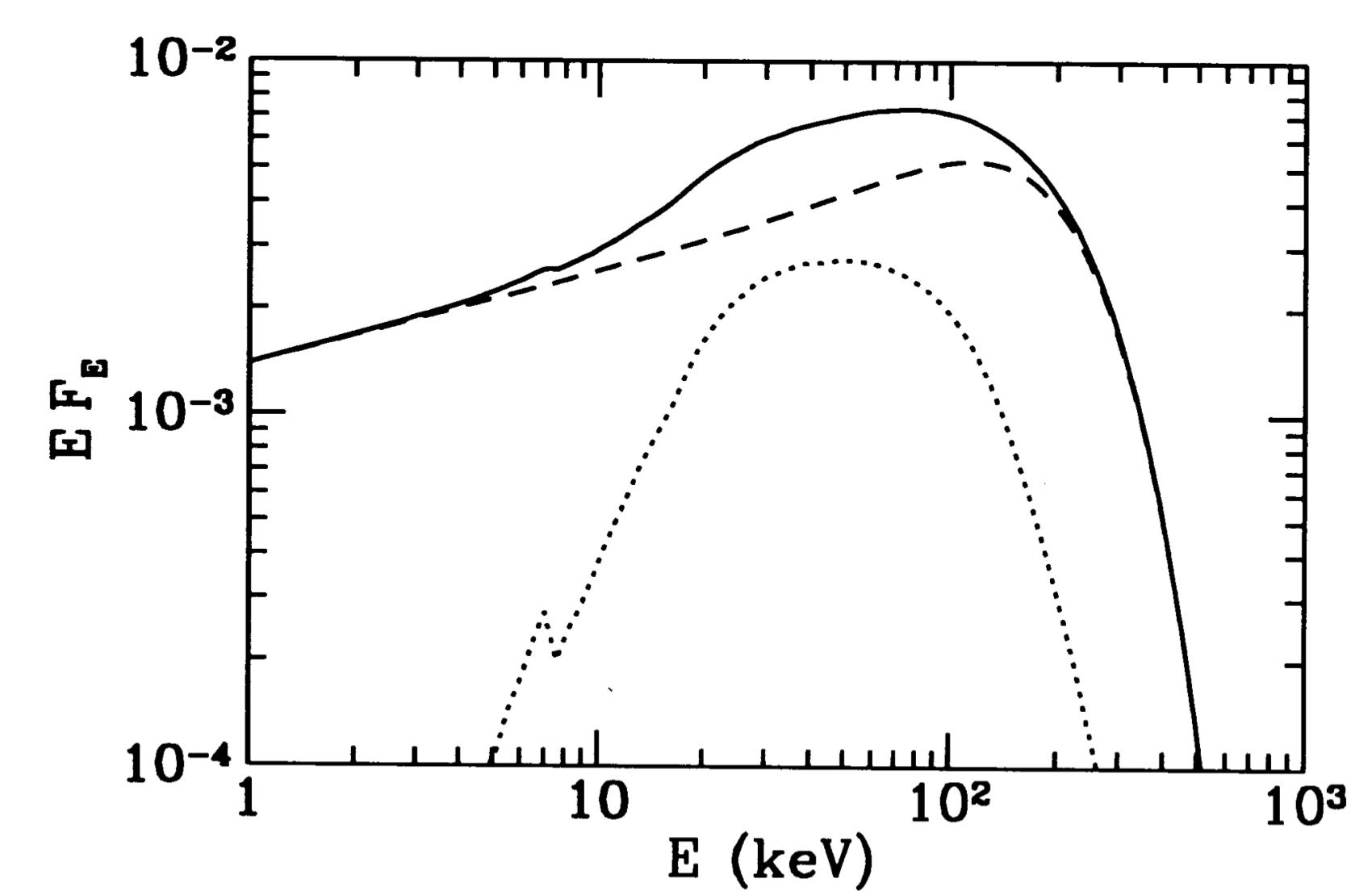}
    \caption{Typical AGN SED produced by the comptonization of UV-optical photons in the corona. Adapted from \cite{zdziarski93}.}
    \label{fig:agn}
\end{figure*}

\begin{figure*}[!ht]
    \centering
    \includegraphics[scale=0.7]{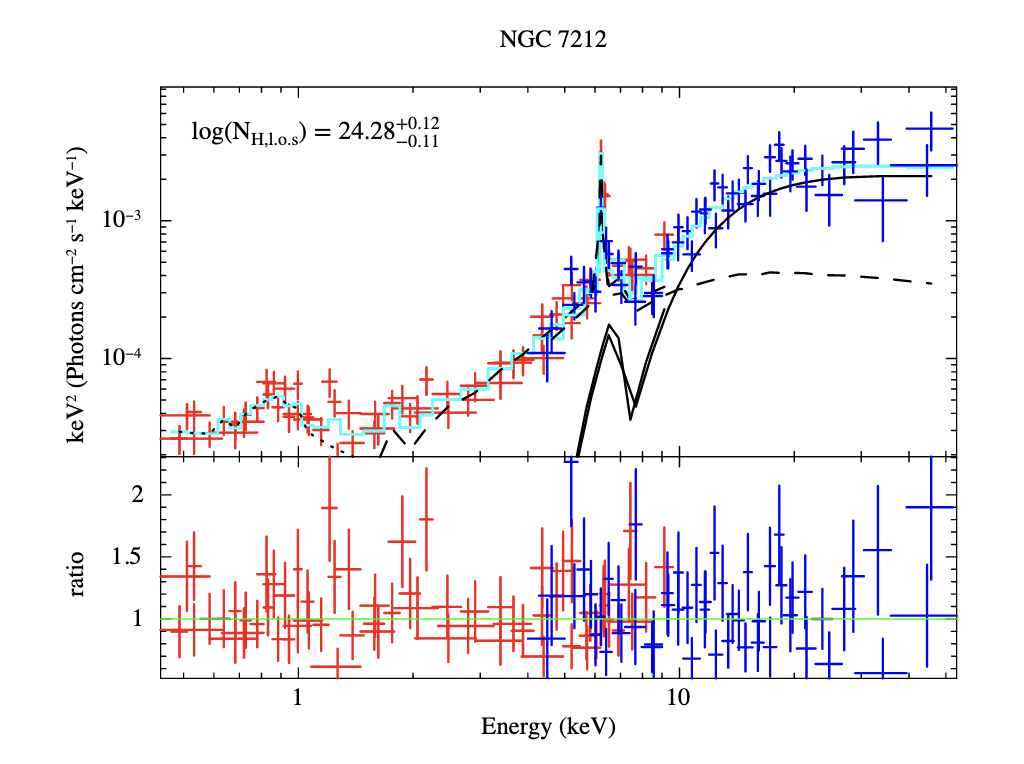}
    \caption{Example of an SED of an obscured AGN. Note how the intrinsic component (solid line) is completely absorbed below $\sim$5\,keV. The dashed line is the reprocess emission by the torus. Figure adapted from \cite{zhao2020}.}
    \label{fig:obs_agn}
\end{figure*}

The optical and UV emission from the inner regions of accretion disks surrounding SMBHs gets upscatterered to the X-ray and MeV band via the interaction of those photons with the hot electrons in the corona, a compact, magnetized region above the disk \citep{haardt93,nandra94}. This produces the typical spectrum of AGN, which is characterized by a power law with exponential cut-off \citep{zdziarski88, zdziarski93}. The photon index of the power law is typically $\approx 1.8$, while the cut-off is around $200-300$\,keV \citep[e.g.][and see Figure~\ref{fig:agn}]{rothschild83,balokovic2020} The cut-off arises because of the thermal distribution of the energies of the electrons in the corona \citep[e.g.][]{poutanen96,petrucci2001}. In the current understanding, the AGN emission is bright in the 1-300\,keV band (and brightest around 10-50\,keV) and becomes faint in the sub-MeV regime (see Figure~\ref{fig:agn}). The spectrum of this intrinsic, nuclear emission is modified by the interaction  of the intrinsic radiation with the matter around the SMBH. In particular, if our line of sight passes through the torus, then this intrinsic emission can be absorbed and Compton-scattered in the interaction with the matter in the torus. This produces the typical spectrum of an obscured AGN, whose emission can be nearly completely attenuated below 10\,keV (see Figure~\ref{fig:obs_agn}).

\begin{figure*}[!ht]
    \centering
    \includegraphics[scale=0.7]{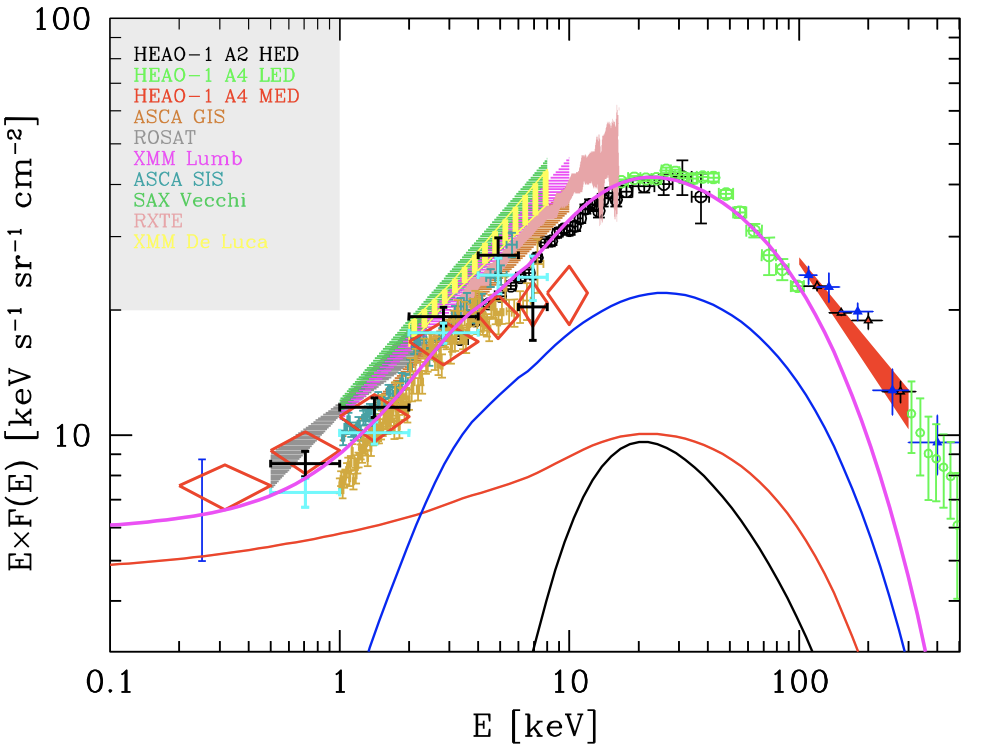}
    \caption{The intensity spectrum of the cosmic X-ray background. Note how the emission from AGN can reproduce the CXB intensity across most of the energy range except at $\gtrsim$100\,keV. Figure adapted from \cite{gilli2007}.}
    \label{fig:cxb}
\end{figure*}

The population of obscured and unobscured AGN is able to reproduce the shape of the Cosmic X-ray Background (CXB) between 1 and 200\,keV showing that our understanding of their emission is very accurate and that the CXB is mostly generated by AGN \citep{gilli2007,treister2009}. In the current paradigm, however, the AGN emission falls short of the CXB emission at $\gtrsim$100\,keV because of the cut-off in the AGN spectrum (see Figure~\ref{fig:cxb}). Here we should note that such cut-off, which falls in the sub-MeV band, is typically beyond the energy range of current and past instruments and has been detected with certainty only in a few objects \citep[see e.g.][]{lanzuisi2019}. The similarity of the CXB spectrum and the integrated one of AGN across the entire 1-200\,keV band and the fact that the CXB spectrum has a power-law shape in the 200-500\,keV band
led some authors to speculate that the corona may also contain a small fraction of non-thermal electrons which can produce a power-law tail in the spectra of AGN that may extend to the MeV band \citep{inoue2008}. 
In their modeling, \cite{inoue2008} find that 3.5\,\% of the total electron energy is carried by non-thermal electrons. These non-thermal electrons may be accelerated via magnetic reconnection.
Interestingly, claims for the existence of both power-law tails in galactic binaries and presence of  non-thermal electrons in AGN have been reported in the literature \citep[see e.g.][]{cadol2006,inoue2018}. Recently a study, using e-Merlin, VLA and ALMA, of the corona of NGC 1068 reports the presence of non-thermal electrons with an energy density of 10\,\% of the total electron energy density \citep{mutie2025}.
The MeV band, thus, emerges as a particularly well-suited energy range
 to study the transition between thermal and non-thermal processes in AGN.

\begin{figure*}[!ht]
    \centering
    \includegraphics[scale=0.5]{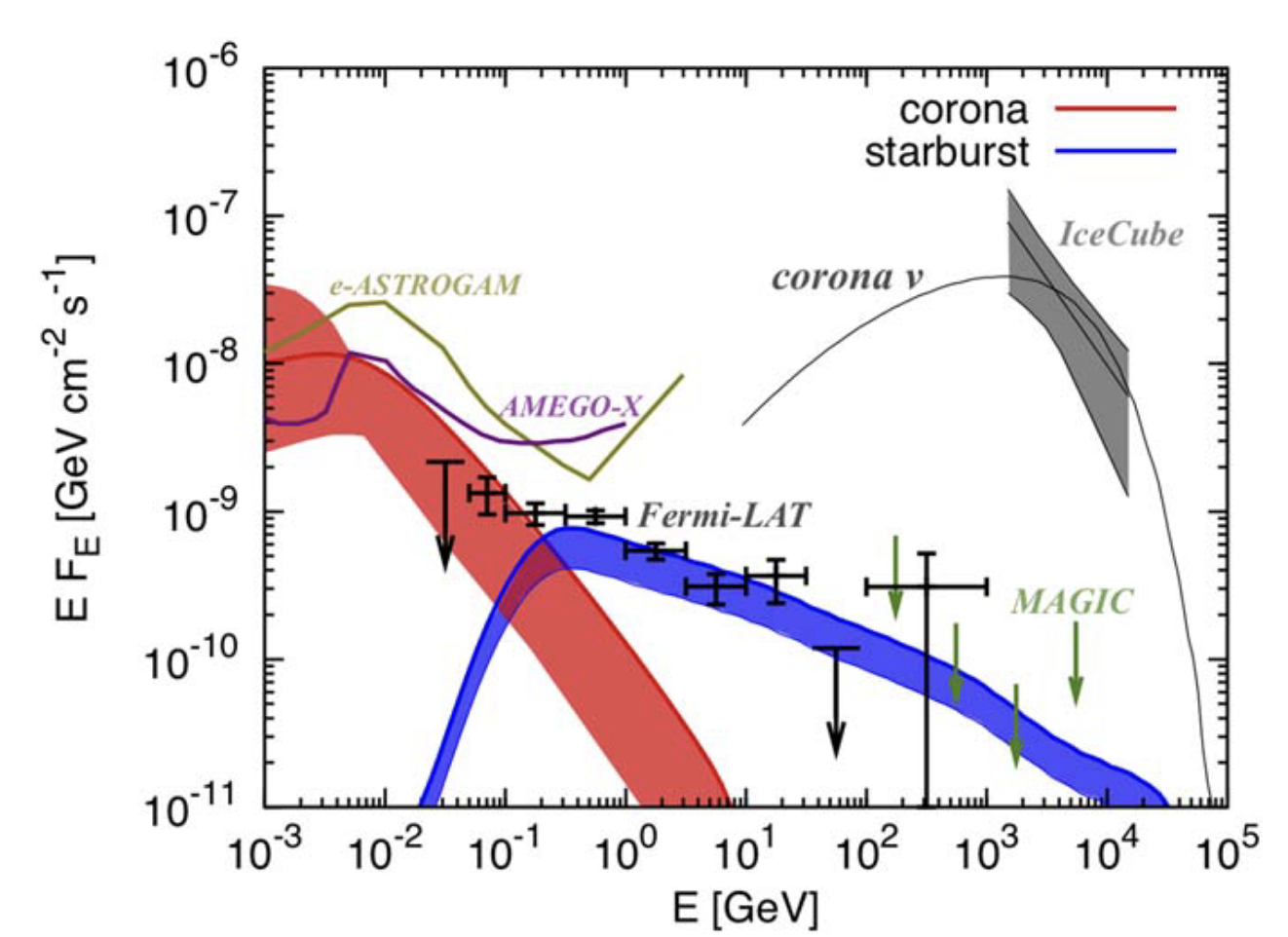}
    \caption{Gamma-ray (black data points) and neutrino (gray bow-tie) SED of NGC 1068 reproduced by a hadronic corona model plus hadronic star formation emission.  Figure adapted from \cite{ajello2023}.}
    \label{fig:ngc1068}
\end{figure*}

Moreover, the IceCube detection of high-energy neutrinos in the direction of NGC 1068 (a prototypical nearby AGN) suggests the presence of a hadronic source in NGC 1068 \citep{icecube_ngc1068}. Because neutrinos and $\gamma$-rays are generated in hadronic interaction with similar energies and in similar numbers, the non-detection of gamma rays from NGC 1068 implies that those gamma rays must be absorbed before they can leave the host \citep{murase2020}. The only place where this can happen (in the entire galaxy) is the corona, where the abundance of X-ray radiation provides the optimal environment to both neutrino generation and gamma-ray absorption. The gamma rays are then reprocessed in the MeV band where they can finally leave the source \citep[see e.g. Figure~\ref{fig:ngc1068} and][]{ajello2023}. While it is not clear yet, if all AGN are neutrino emitters, those that are must be brightest in the MeV band. Observations of these sources at MeV energies uniquely probe the distance of the emission region from the SMBH and the emission process.

\subsection{Relativistic jets}

\subsubsection{How to trace relativistic jet presence across cosmic time}

The presence of relativistic jets in AGN has historically been associated with a strong radio-loudness parameter. $R=F_{\rm radio}/F_{\rm opt}>10$ is the standard threshold\footnote{
The $R$ parameter  traditionally refers to the ratio between the rest-frame radio flux at $\nu\sim5$ GHz and the rest-frame optical flux in the blue band $B$, but some works have used adjacent frequencies and bands to define it (e.g.\ 1.4 instead of 5 GHz). For this reason, we refer generally to $F_{\rm radio}$ and $F_{\rm opt}$ in the text. 
} above which the radio flux emitted via isotropic (extended lobes) or relativistically beamed synchrotron emission (beamed jet) dominates over the nuclear optical-UV emitted by the SMBH accretion flow \citep{fanaroffriley74}. Approximately $10\%$ of the AGN population is radio-loud, suggesting an occurrence of roughly one jetted source every 10 AGN. 

In the last decade, successful exploration of the high-$z$ quasar population allowed the study of their radio-loudness fraction, hoping to recover the jet occurrence in AGN in the first Gigayears of the Universe. Interestingly, the number of radio-loud sources is consistently one tenth of the overall known AGN population at all redshifts \citep{liu21}, but the radio fluxes are notably obtained with different approaches: at low redshift, the FIRST survey \citep{white97} provides a good and uniform coverage of the northern sky down to a sensitivity limit of 1 mJy. This sensitivity does not allow for a complete survey of sources at higher redshift, and the radio detection must be provided via dedicated observations. This can lead to inconsistencies, and an independent approach has been explored.

Blazars can be extremely efficient tracers of the jetted AGN population. They are defined as jetted AGN with their jets aligned along our line-of-sight. If a stricter definition is considered, i.e.\ if only those sources whose viewing angle $\theta_{\rm v}$ is smaller than the jet beaming angle $\theta_{\rm b}\simeq1/\Gamma$ (where $\Gamma$ is the bulk Lorentz factor of the jet emitting region) are defined as blazars, then each blazar detection infers the presence of $2\Gamma^2$ analogous jetted sources with their jets directed randomly in the sky. Given that $\Gamma$ values typically range between 10 and 15 \citep{ghisellini2014}, following simple geometric assumptions every blazar must trace the presence of few hundreds (100-450) jetted AGN with the same SMBH and jet features (e.g.\ mass, accretion rate, jet power, energetics and particle composition), and in the same redshift bin \citep{ghisellini10,volonteri11}. Clearly this is a very efficient way to perform population studies, provided that all-sky blazar catalogs are available. 

The typical smoking gun to classify a blazar is its very high-energy emission, i.e.\ the AGN association to a bright source in the $>100$ MeV energy range. {\it Swift}/BAT and {\it Fermi}-LAT provide all-sky coverage in the hard X-ray and $\gamma$-ray energy ranges, allowing for a complete compilation of blazars up to $z\sim3.5-4$. At redshifts larger than 3, the majority of AGN are quasars with large masses. This is likely a selection effect, since only the brightest sources can be identified, and emission lines are crucial to derive their redshift. To study the jet occurrence in AGN across cosmic time is therefore useful to focus on $M_{\rm BH}\geq10^9M_\odot$ showing big blue bumps and broad emission lines, i.e.\ quasars. With this approach, the populations at different redshifts can uniformly be compared. Following this approach, \cite{volonteri11} and \cite{ghisellini10} found hints that the jet distribution in extremely massive, high-$z$ AGN is different than the average, local $\sim10\%$ occurrence. 
In fact, comparing optical quasar luminosity functions with the high-energy ones, it is clear that at $z>3.5$ jets are more frequent than at lower redshift. With the help of systematic blazar search and identification at $z>4$ this tendency was confirmed: most (if not all) $>10^9M_\odot$ black holes in the first 1.5 Gyrs of the Universe are hosted in jetted AGN \citep{belladitta20,sbarrato21,sbarrato22}.

\subsubsection{Evolution of FSRQ across cosmic time: MeV blazars}

High-$z$ blazars are currently among the brightest blazars observed, and are typically characterized by broad-band Spectral Energy Distributions (SEDs) with strong Compton Dominance (i.e.\ Inverse Compton hump significantly dominating over the synchrotron one), accompanied by a low frequency synchrotron peak. 
They are easily classifiable as archetypal Flat Spectrum Radio Quasars (FSRQs), thanks to their powerful and red SEDs, often representing the brightest among these sources. 
Their Inverse Compton component can peak at frequencies as low as the MeV range (2-100 MeV), while less powerful blazars observed at lower redshifts peak in the $\gamma$-rays: they have many common features with the lately defined ``MeV blazars". 
These extreme and powerful blazars are thought to be responsible for almost the entirety of the cosmic MeV background \citep{marcotulli22}. 
{\it Swift}/BAT has been key to identify and study such sources, thanks to its ability in detecting their hard X-ray emission, and well describing the peak or the rising part of their Inverse Compton component. {\it Fermi}/LAT is also a useful instrument in studying the details of their high-energy SEDs: in its lower-energy range, it can detect the $\gamma$-ray emitted right after the SED peak. 
However, only a fraction of the known MeV blazars is detected by both instruments, allowing for a precise reconstruction of their jet emission and energetics. 

BAT-detected blazars have been thoroughly studied and compared with $\gamma$-ray detected blazars \citep{marcotulli22}, 
finding that they appear to have a significantly brighter inverse Compton component and smaller peak frequencies. If they are divided in luminosity bins, though, the expected anticorrelation between jet power and peak frequency is not observed. 
Broad-band SED studies also suggest that the bulk Lorentz factor of the emitting region might be in the range $\Gamma\sim8-10$, slightly smaller than what observed among {\it Fermi}-detected blazars \citep[$\sim13-15$,][]{ghisellini2014}. 
Clearly, {\it Swift}/BAT is sampling an extreme of the jet power distribution, but its sensitivity does not allow to deepen our knowledge of the so-called MeV blazars, lacking the possibility of exploring a potential lower luminosity ``counterpart" of such sources. 
A sensitive instrument in the MeV range would be crucial to explore how these sources connect with the overall blazar population, both by verifying their existence at different redshifts and probing lower luminosity jets that are not yet observed at high energies. 
In particular, if the different behaviour of the inverse Compton component will be confirmed between MeV and $\gamma$-ray blazars, it would suggest different energetics of the particles involved in such emission. 

\begin{figure*}[!ht]
    \centering
    \includegraphics[width=0.8\textwidth]{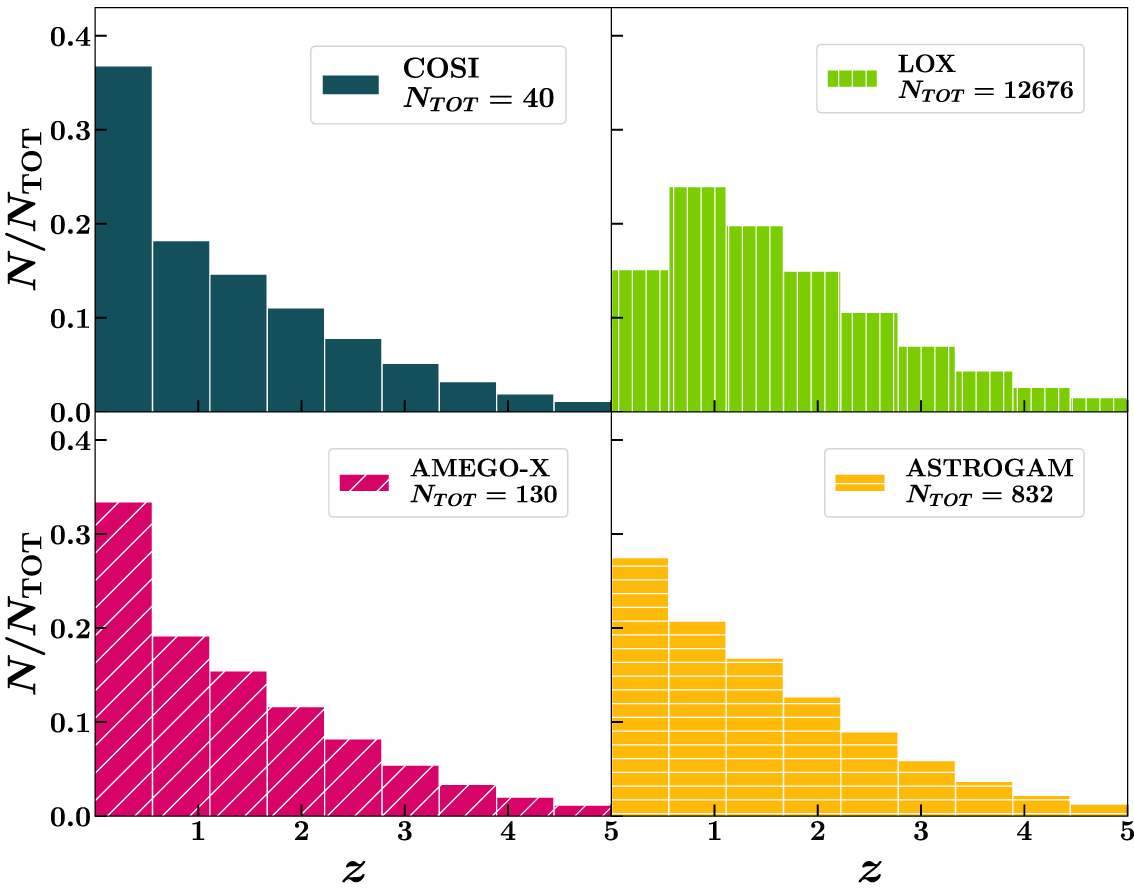}
    \caption{Fraction of MeV sources expected to be detected by the labelled missions, 
             as a function of redshift. All the planned missions will detect a significant amount of sources 
             even at very high redshift, allowing for a deeper understanding of the most powerful 
             relativistic jets across cosmic time.
             From \cite{marcotulli22}.
             }
    \label{fig:mev-blaz-numbers}
\end{figure*}

The evolution in cosmic time of powerful blazars is also of particular interest because it seems that BAT- and LAT-detected blazars show discrepancies in their space density distributions: $\gamma$-ray blazars peak at $z\sim1.6$, while MeV sources peak significantly later ($z\sim4.3$). 
It has been suggested that this significant discrepancy may be ascribed to two different evolutionary channels: BAT- and LAT-detected blazars might belong to the same family of sources, but underwent different paths in their evolution.
However, it must be noted that {\it Fermi}-LAT has a shallower sensitivity, and therefore its horizon is much smaller than {\it Swift}/BAT: the derived density distribution might be biased. 
To explore whether these two groups are actually different, the MeV band is instrumental: other than a more complete compilation of MeV blazars, it will probe the inverse Compton component of higher-redshift counterparts of $\gamma$-ray blazars, that {\it Fermi}-LAT is unable to detect, confirming (or disproving) whether there  are two slightly different population of powerful FSRQs evolving on different cosmic timescales.
A preliminary study of the expected blazar population peaking in the MeV was performed in \cite{marcotulli22}, who estimated that existing and forthcoming X-ray and soft $\gamma$-ray missions will be able to detect new MeV blazars in a wide redshift range (see Figure \ref{fig:mev-blaz-numbers}): deep MeV observations will be instrumental to characterize this population.

\begin{figure*}[!ht]
    \centering
    \includegraphics[width=\textwidth]{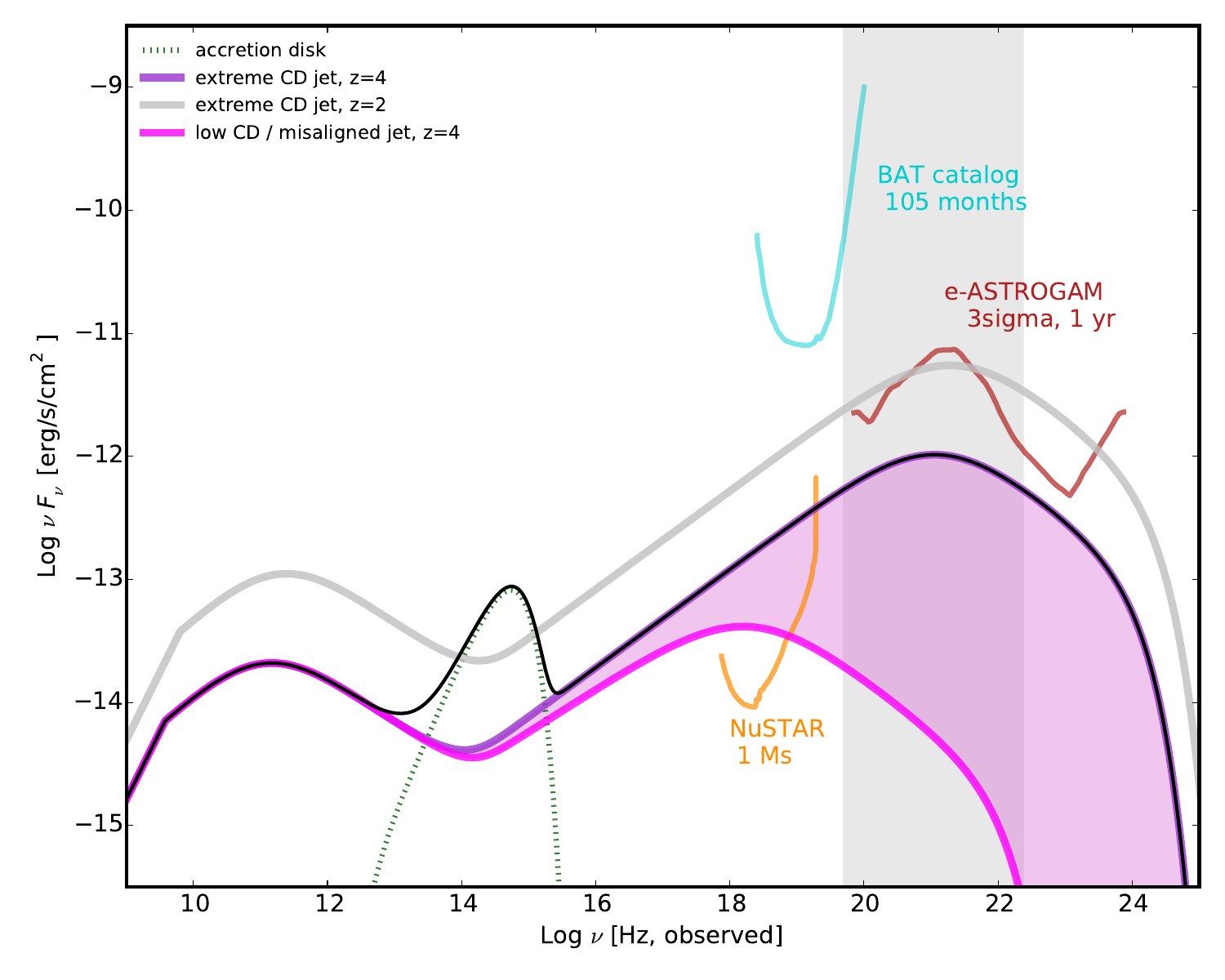}
    \caption{Broad band SEDs of a typical high-$z$ blazar, 
             compared with the sensitivity curves of {\it Swift}/BAT and {\it NuSTAR}. The grey shaded area shows the MeV frequency range, where it would be possible to observe the peak emission of the source. The sensitivity proposed for e-ASTROGAM \citep{2018JHEAp..19....1D} is also shown.
             }
    \label{fig:blazar_SED_curves}
\end{figure*}

\subsubsection{Jets in the highest redshift blazars}

The first high-$z$ blazars serendipitously identified in the last 20 years were remarkably similar to the ones observed in the lower redshift Universe, i.e.\ FSRQs with strong Compton Dominance, prominent big blue bumps and a bright, flat radio spectrum. 
Once the observed population started to increase, the discovered sources started to show more diversity in their multi-band features. 

The systematic classification approach followed at $z>4$ relies on mainly two characteristics: a flat radio spectrum with a compact radio emission, and/or a bright and hard X-ray spectrum. Different groups followed different approaches, relying only on a specific band or favoring a hybrid approach \citep{caccianiga19,frey18,sbarrato13}. Few sources started to show some inconsistencies in their multi-wavelength behaviours, triggering particular interest in the community. 

Some radio bright quasars were classified as bona fide blazars thanks to dedicated soft X-ray observation and broad-band SED modeling, but showed features of misaligned, extended jets in follow-up high-resolution radio observations \citep[see e.g. J1420+1205 and J2220+0025 in][]{sbarrato15,cao17}. 
These inconsistencies lead to various possibilities, that cannot be evaluated and disentangled with the available observations and instruments (as shown in Figure \ref{fig:blazar_SED_curves}).
\begin{itemize}
    \item The X-ray emitting region is located at about few tenths of a parsec from the central engine, while the extended emission can be located few hundreds of parsec away. {\it The jet orientation might have changed between the two regions}. Changes in density of the surrounding medium might bend the propagation direction of the jet, or a chaotic accretion history might induce changes in the SMBH angular momentum and thus in the jet direction close to its base. 
    \item The X-ray spectrum is not produced by the relativistically beamed jet, but is instead a product of {\it upscattered CMB photons}: at $z>3$, the CMB energy density\footnote{The CMB energy density scales as $U_{\rm CMB}\propto(1+z)^4$, and is therefore significant at high redshift. } becomes competitive with the comoving energy density of the magnetic fields permeating the extended lobes, and typically responsible for the synchrotron emission and its strong radio luminosity \citep{tavecchio00}. The leptons populating the extended jets might interact more easily with CMB photons via inverse compton processes, producing an extended X-ray emission with a somewhat hard X-ray spectrum. 
    \item Finally, it cannot be excluded that the {\it jet emission processes are significantly different} at higher redshift than in the local Universe, e.g.\ the jet bulk Lorentz factors might be smaller, allowing for X-ray detection from the relativistic emitting region even at much larger viewing angles. This is less likely, because it would lead to jet powers significantly out of equipartition, that is less expected \citep{sbarrato22}.
\end{itemize}

The first two options are the most likely, and thus the most interesting to be investigated. 
Currently, no observations at frequencies larger than $\sim8-10$ keV are available for the debated sources, thus preventing us from observing the harder X-ray and MeV profile: no instrument is currently able to map that energy range. 
Hard X-ray telescopes such as {\it NuSTAR} might be able to trace the low frequency edge of the SED different behaviours, but not enough to reliably discriminate. 
The MeV range would instead allow to discriminate between a standard blazar-like X-ray emission (that would favour the bending hypothesis) and a typical IC/CMB emission profile, that would instead peak at much smaller frequency, accompanied by a significantly smaller IC luminosity. 
The required sensitivity for these ideal observations is anyway very challenging even for the currently discussed MeV-focused instrument: developing new technological concepts and telescope projects to investigate the MeV window is key to advance our knowledge of the earliest jets forming in our Universe.

\section{Jets and Outflows Physics: more than photons}

\subsection{Neutrinos and their implication on jet emission}

Astrophysical high-energy ($>$TeV) neutrinos are predicted to arise from hadronic interactions occurring in or near cosmic-ray accelerators. The IceCube Observatory's detection of a diffuse astrophysical neutrino flux provided the first evidence for hadronic high-energy cosmic-ray accelerators beyond our galaxy. The isotropic distribution of these neutrino events across the sky suggests a significant extragalactic origin~\citep{2013Sci...342E...1I}.  Although no individual neutrino point source has been identified with high confidence to date, there is supporting evidence pointing to active galactic nuclei, particularly blazars, as potential sources. In 2017, a high-energy neutrino event, IceCube-170922A -- of high probability of being astrophysical --, was detected from the direction of the $\gamma$-ray blazar TXS 0506+056~\citep{2017GCN.21916....1K}. 

The arrival of the neutrino coincided with a period of enhanced $\gamma$-ray emission from TXS 0506+056, with the likelihood of this association being by chance estimated at the 3 sigma level \citep{2018Sci...361.1378I}. In blazars, gamma rays can be produced by leptonic or hadronic interactions. The association of neutrinos with blazars indicates the presence of relativistic hadrons in blazar jets. High-energy neutrino production occurs as a result of the decay of charged pions produced in the interactions of high-energy protons with matter ($pp$) (hadronic) or with photon fields ($p\gamma$) (photo-hadronic). Gamma rays are produced when neutral pions ($\pi_0$), created alongside the charged pions, decay into two photons. 

\begin{figure}
    \centering
     \includegraphics[width=0.9\textwidth]{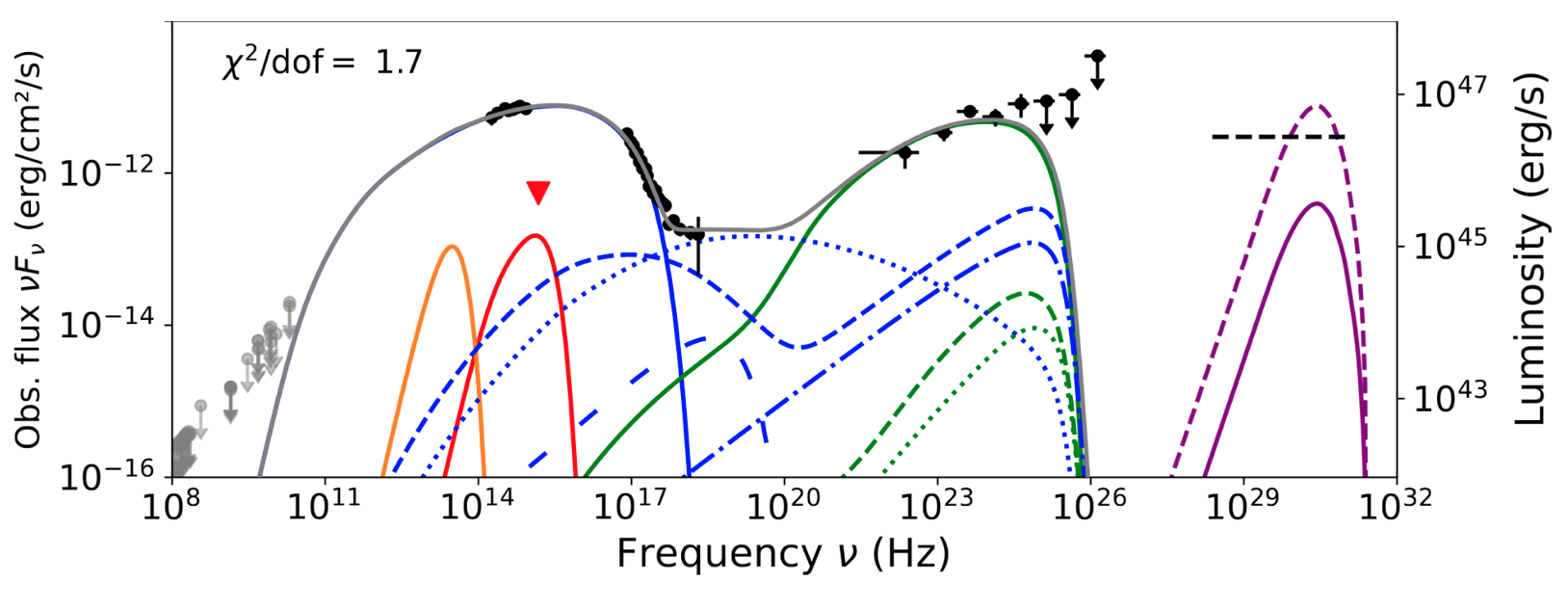}
    \caption{Lepto-hadronic modeling of the SED of the blazar 5BZB~J0630$-$2406 in the observer frame \citep{Buson_j0630:2024,J0630}. Archival data are shown by the gray solid dots. The red downward triangle represents the accretion-disk limit \citep{Azzollini:2024}.
    The lepto-hadronic modeling (gray line) adequately describes the SED, built with the quasi-simultaneous optical/X-ray data from Neil Gehrels $Swift$ Observatory, XMM-Newton and $NuSTAR$, and the $\gamma$-ray data from $Fermi$-LAT (black points).
    Among other components, the dashed lines highlight hadronic processes (Bethe-Heitler and photopion productions), becoming important mostly in the soft-to-hard X-ray band, up to the MeV range.
    The predicted neutrino flux emerges at the highest energies (purple line) with associated uncertainties from Poisson statistic (dashed line, $3\sigma$), and is at reach of the 7-yr IceCube Observatory sensitivity \citep{IceCube7y:2017}.
    }
    \label{fig:J0630}
\end{figure}

Blazars are highly variable in nature, and flares from individual gamma-ray blazars provide promising opportunities for the identification of neutrino emitters \citep[e.g.,][]{2018ApJ...865..124M}. Following the first compelling association of a blazar with an IceCube neutrino, a few hundreds of similar high-energy neutrino events have been recorded by the IceCube Observatory. However, despite studies tackling the neutrino/$\gamma$-ray link with increased statistics, it remains inconclusive on whether, in blazars, the bulk of observed gamma rays can be directly associated with neutrino emission \citep{Garrappa}.
Theoretical models predict that the $\gamma$-rays cospatially produced with the neutrino should cascade down to lower energies. As a result, one may expect this emission to emerge at the lower energies, in the X-rays / MeV band. A recent study focused on a candidate neutrino-emitter blazar \citep{journey,Buson_erratum:2022,Buson_2023}, i.e. 5BZB~J0630$-$2406, suggests that the hadronic imprint may have been already serendipitously observed in the blazar SED \citep{J0630}.
Based on the theoretical predictions presented in the study, the blazar is capable of producing neutrinos within the reach of the IceCube detector, and can plausibly contribute to the anisotropy observed in the distribution of IceCube events observed in its direction. Its SED can be adequately modeled via both purely leptonic and mixed lepto-hadronic scenario, except for one epoch during which quasi-simultaneous Neil Gehrels {\it Swift} Observatory, XMM-Newton and $NuSTAR$, and $Fermi$-LAT observations are available. During this epoch, the evidence for a break in the spectral shape of the X-ray emission suggests the presence of an additional component. The latter is  challenging to be reproduced with a pure leptonic model (see Figure \ref{fig:J0630}). On the other hand, within the hadronic scenario, such component is naturally explained by the processes related to very energetic protons (e.g. Bethe-Heitler). This potential hadronically-powered emission is predicted to become dominant in the SED from soft-X-ray to MeV band, highlighting the largely unexplored MeV-band as particularly promising for identifying hadronic signatures in the blazar's SED.

\begin{figure*}[!ht]
    \centering
    \includegraphics[width=\textwidth]{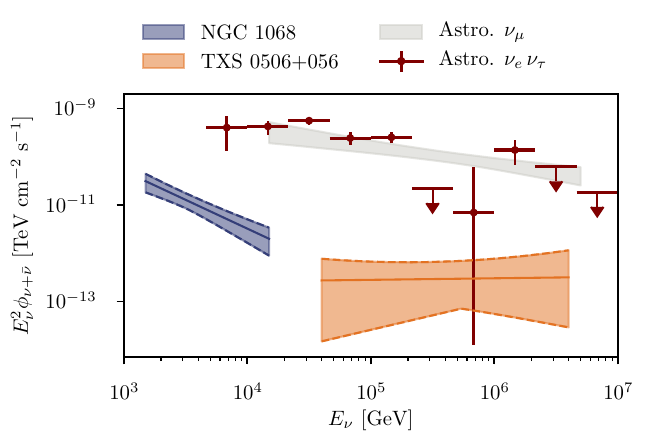}
    \caption{Point-source neutrino fluxes for NGC 1068 and TXS 0506+056 measured by IceCube in comparison with the total diffuse astrophysical neutrino flux. Figure from~\cite{icecube_ngc1068}. 
    }
    \label{fig:Icecube_NGC1068}
\end{figure*}

As noted earlier in \S\ref{subsect:seyferts}, the IceCube Collaboration reported an encouraging evidence for TeV neutrino emission from the nearby active galaxy NGC 1068, classified as a Seyfert galaxy~\citep{icecube_ngc1068}. Seyfert galaxies belong to the class of non-jetted AGN, unlike blazars, and the origin and nature of their electromagnetic emission is different. While blazars emit a large fraction of their energy through non-thermal processes, such as relativistic jets, the accretion disk around SMBH play a significant role in the emission in non-jetted AGN, with more thermal emission, along with other components, such as the corona.  Figure~\ref{fig:Icecube_NGC1068} shows the measured neutrino fluxes from TXS~0506$+$056 and NGC 1068 as a function of energy, in comparison to the diffuse neutrino flux. While both sources belong to the class of AGN, their neutrino spectral properties are quite different. This suggests that the emission mechanisms are not the same, pointing to a population of neutrino sources whose accompanying gamma-ray emission primarily emerges in the MeV band (see Section \ref{subsect:seyferts}).

\subsection{Jet composition}

Besides the role played in the multimessenger context, the MeV band is also instrumental to explore several issues related to the physics of relativistic jets.

A key feature of the MeV band is that the most powerful FSRQ emit most of their power in this energy range. A measure of the emitted power is therefore a straightforward, but essential, measurement made possible by MeV detectors. In more detail, also the precise frequency of the peak of the high-energy bump is a key information for jet modelling, since it directly informs us about the nature of the seed photons for the external inverse Compton process (that can be dominated either by the emission from the BLR or from the IR dusty torus) and thus the location of the emitting region \citep{sikora2002}. Moreover, the precise location of the peak is one of the critical parameters we need to constrain the emission models and, in turn, key parameters of the jet such as its power \citep{ghisellini2014} and magnetization \citep{janiak2015}. Both quantities are at the focus of the current debate on acceleration and collimation processes of AGN jets (thought to be powered by the extraction of the rotational power of the central supermassive BH or of the accretion disk, through the Blandford-Znajek and the Blandford-Payne mechanisms). The actual magnetization of the jet is also essential in determining the main mechanism behind the acceleration of particle at ultrarelativistic energies, i.e. shocks (efficient only for low magnetized flows), magnetic reconnection or turbulence \citep[e.g.,][]{matthews2020}. 

Besides being sensitive to jet magnetization, the MeV emission from jets carries unique imprints of the matter composition of the outflowing plasma. In particular, if protons are present in the jet and are accelerated at ultrarelativistic energies, they could provide (besides the neutrino output) unique radiative signatures that can be exploited to probe the composition. In fact, hadronic models predict several emission components (produced by electrons/pairs from Bethe-Heitler or electromagnetic cascades) contributing between 100 keV and tens of MeV (see the previous section). 

The polarization channel is perhaps the most powerful tool to probe the emission mechanisms at work in the gamma-ray band and, indirectly, the underlying emitting particles (hadrons or leptons). In particular, while the high-energy component in leptonic model (produced through IC scattering) is expected to display a low level of polarization (each scattering halves the net polarization of the seed radiation), hadronic scenarios generally predict highly polarized radiation. In fact, in latter models the gamma-ray emission receives a substantial contribution from synchrotron radiation produced either directly by high-energy protons or by $e^\pm$ pairs byproduct of photomeson production and subsequent cascades. The degree of polarization can be as large as 70-80\% in the MeV band \citep{zhang2013} for uniform fields. However, the observed polarization in the optical band and, more recently, in the X-ray band by the {\it IXPE} satellite, suggests the existence of some degree of turbulence (i.e. disordered fields) or magnetic fields with multiple components (i.e. comparable toroidal and poloidal components), which substantially decreases the net polarization. However, as demonstrated in \cite{Paliya2018}, even in this case hadronic scenarios could provide a degree of polarization similar to that in the X-ray band (as large as 20-30\%), much larger than that predicted for leptonic models in which gamma rays are produced through IC scattering. COSI will be the first MeV instrument  potentially accessing this observational window, but it will have a limited sensitivity, thus making difficult to obtain conclusive results.

Another tool potentially able to discriminate between hadronic and leptonic emission is the multiwavelength variability. Since in hadronic scenarios one decouples the particles mostly contributing to the low and the high-energy peaks, MW lightcurves could display different behaviors at different frequencies (while in leptonic models for FSRQ we expect almost strictly similar variability in e.g. MeV and soft X-rays or optical). In extreme conditions even orphan flares (i.e. variability of the high-energy band without counterpart at lower energies) are possible, see e.g. \cite{diltz2016}. Clearly, a high-sensitivity monitoring of the gamma-ray sky, accompanied by dense monitoring at lower frequencies, is instrumental to exploit this technique.

A long-standing issue in the study of jets is the possible presence (or prevalence) of $e^+$-$e^-$ pairs in the outflowing plasma. Although several arguments help to constrain the pair to proton number ratio of the order of a few tens at most \citep[e.g.,][]{ghisellini2012}, some recent studies argues for a dynamically dominant role of pairs e.g., \cite{zdz2022}. The MeV band can play a crucial role in this issue, since sensitive instruments could be able to directly detect the emission coming from the expected pair annihilations. In particular, \cite{marscher2007} proposed that a narrow annihilation line could be directly observable when jets strongly interact with the external gas, allowing the efficient mixing of the jet material with the external environment. On the other hand, \cite{ghisellini2012} showed that efficient annihilation could occur close to the SMBH, when the jet is sill in the acceleration phase. This would provide a powerful broad MeV component observable in misaligned jets (i.e. radiogalaxies).

\section{Summary and Comments}

This work explores the potential MeV signatures from the environments of SMBHs, starting from our relatively inactive Galactic Centre and its high-energy emission, then moving through the lesser-understood nuclear structure of AGN, the X-ray Corona, up to the most extreme relativistic jets, their composition, and evolution.
The MeV emission from all these different structures reveals a strong common link: the most pressing uncertainties across these environments concern their particle composition and  emission properties. 
\begin{itemize}
    \item[-] 
    The particle population around the Galactic Centre can be characterized based on its dominant energy loss mechanism. The expected $\gamma$-ray flux from the so-called {\it Fermi} Bubbles would differ most significantly at energies below 50 MeV between the much brighter leptonic scenario and the likely 5 times dimmer hadronic scenario. The latter would most likely suggest quasi-stationary power injections due to mini-bursts from SgrA$^*$, while the leptonic scenario is supposed to be linked to a starburst dominated hypothesis. 
     \item[-] 
    Although AGN dominate the cosmic X-ray background, the hard-X- to soft-$\gamma$-ray emission from these sources remains surprisingly poorly understood. Non-thermal processes are certainly required to justify the X-ray brightness of this population, but the specifics of how and at which frequencies the nuclear emission transitions from thermal to non-thermal remain widely open questions. The possible production of high-energy neutrinos in AGN without prominent relativistic jets suggests the presence of hadronic sources in their nuclei. MeV photons are a relevant product of neutrino emission, and their detection would provide insights into the geometry and energetics of the expected non-thermal emitting regions in AGN (i.e.\ the X-ray Corona). 
    \item[-]
    When relativistic jets are instead present and aligned closely to our line-of-sight, non-thermal emission dominates at all wavelengths. Neutrinos from jetted AGN have revived the hypothesis of a significant hadronic component in relativistic jets, that could be probed at MeV energies, even in this relativistically beamed form. Strong magnetic fields, which typically permeate relativistic jets, would also be responsible for high polarization degrees in hadronic scenarios, dominated by synchrotron emission from hadrons themselves or from leptonic by-products. 
    In contrast, leptonic-dominated emission processes are not expected to show significant polarization at MeV energies. Thus, studying polarized emission in this energy range could provide robust clues for understanding relativistic jets composition and their emission processes.
    \item[-]
    Another open question regarding relativistic jets in AGN is their evolution across cosmic time. The most powerful blazars, visible up to very high redshift, exhibit peculiar emitting features compared to their less powerful and more local counterparts. The jet's extension, structure and its interaction with the CMB strongly affect the high-energy emission. The $\gamma$-ray emission of these jets shows a steep decline, making their detection at $z>3$ extremely difficult with current facilities.
    Opening the MeV observing window will help us understand how such extreme sources evolve, if they follow the same patterns as their lower-counterparts counterparts, and ultimately how the accretion-ejection relation evolves with cosmic time.
\end{itemize}

The MeV energy range offers an exceptionally powerful window for understanding the physics, composition and emission processes of matter in the immediate surroundings of SMBHs. 
Whether during their active phases or quiescent periods, SMBHs significantly influence their environment, producing some of the most powerful phenomena of the high-energy sky. 
An MeV telescope with polarimetric capabilities will undoubtedly provide new insights into how particles accelerate, interact, and radiate in such extreme environments.

\section*{Declarations}

\subsection*{Competing Interests and Fundings}

No specific funding was received to assist with the preparation of this manuscript.
The authors have no competing interests to declare that are relevant to the content of this article.

\def\aj{AJ}%
\def\actaa{Acta Astron.}%
\def\araa{ARA\&A}%
\def\apj{ApJ}%
\def\apjl{ApJ}%
\def\apjs{ApJS}%
\def\ao{Appl.~Opt.}%
\def\apss{Ap\&SS}%
\def\aap{A\&A}%
\def\aapr{A\&A~Rev.}%
\def\aaps{A\&AS}%
\def\azh{AZh}%
\def\baas{BAAS}%
\def\bac{Bull. astr. Inst. Czechosl.}%
\def\caa{Chinese Astron. Astrophys.}%
\def\cjaa{Chinese J. Astron. Astrophys.}%
\def\icarus{Icarus}%
\def\jcap{J. Cosmology Astropart. Phys.}%
\def\jrasc{JRASC}%
\def\mnras{MNRAS}%
\def\memras{MmRAS}%
\def\na{New A}%
\def\nar{New A Rev.}%
\def\pasa{PASA}%
\def\pra{Phys.~Rev.~A}%
\def\prb{Phys.~Rev.~B}%
\def\prc{Phys.~Rev.~C}%
\def\prd{Phys.~Rev.~D}%
\def\pre{Phys.~Rev.~E}%
\def\prl{Phys.~Rev.~Lett.}%
\def\pasp{PASP}%
\def\pasj{PASJ}%
\def\qjras{QJRAS}%
\def\rmxaa{Rev. Mexicana Astron. Astrofis.}%
\def\skytel{S\&T}%
\def\solphys{Sol.~Phys.}%
\def\sovast{Soviet~Ast.}%
\def\ssr{Space~Sci.~Rev.}%
\def\zap{ZAp}%
\def\nat{Nature}%
\def\iaucirc{IAU~Circ.}%
\def\aplett{Astrophys.~Lett.}%
\def\apspr{Astrophys.~Space~Phys.~Res.}%
\def\bain{Bull.~Astron.~Inst.~Netherlands}%
\def\fcp{Fund.~Cosmic~Phys.}%
\def\gca{Geochim.~Cosmochim.~Acta}%
\def\grl{Geophys.~Res.~Lett.}%
\def\jcp{J.~Chem.~Phys.}%
\def\jgr{J.~Geophys.~Res.}%
\def\jqsrt{J.~Quant.~Spec.~Radiat.~Transf.}%
\def\memsai{Mem.~Soc.~Astron.~Italiana}%
\def\nphysa{Nucl.~Phys.~A}%
\def\physrep{Phys.~Rep.}%
\def\physscr{Phys.~Scr}%
\def\planss{Planet.~Space~Sci.}%
\def\procspie{Proc.~SPIE}%
\let\astap=\aap
\let\apjlett=\apjl
\let\apjsupp=\apjs
\let\applopt=\ao
\bibliography{bibliography}

\end{document}